\documentclass[%
 aip,
 amsmath,amssymb,
 reprint,%
]{revtex4-1}

\usepackage{graphicx}% Include figure files
\usepackage{dcolumn}% Align table columns on decimal point
\usepackage{bm}% bold math
\usepackage{xcolor}

\usepackage[utf8]{inputenc}
\usepackage[T1]{fontenc}
\usepackage{mathptmx}
\newcommand{\znse}{\textrm{C\MakeLowercase{r}}$^{2+}$\textrm{:Z\MakeLowercase{n}S\MakeLowercase{e}}}
\begin{document}

\preprint{AIP/123-QED}

\title{Generation of 3 mJ, 44 fs, 2.5 $\mu$m pulses from \\ a single-stage \znse{} amplifier}
% Force line breaks with \\

\author{Yi. Wu}
 \affiliation{CREOL and Department of Physics, University of Central Florida,\\ Orlando, Florida 32816, USA.}%Lines break automatically or can be forced with \\
\author{Fangjie Zhou}%
\affiliation{CREOL and Department of Physics, University of Central Florida,\\ Orlando, Florida 32816, USA.%\\This line break forced with \textbackslash\textbackslash
}%
\author{Esben W. Larsen}
\email[Corresponding Author: ]{elarsen@imperial.ac.uk}
\affiliation{%
Quantum Optics and Laser Science Group, Blackett Laboratory, Imperial College London,\\ London, SW7 2BW, UK.%\\This line break forced% with \\
}%

\author{Fengjiang Zhuang}%
\affiliation{CREOL and Department of Physics, University of Central Florida,\\ Orlando, Florida 32816, USA.}

\author{Yanchun Yin}%
\affiliation{CREOL and Department of Physics, University of Central Florida,\\ Orlando, Florida 32816, USA.}

\author{Zenghu Chang}
\affiliation{CREOL and Department of Physics, University of Central Florida,\\ Orlando, Florida 32816, USA.%\\This line break forced% with \\
}%

\date{\today}% It is always \today, today,

\begin{abstract}
Lasers capable of generating attosecond X-ray pulses in the water window (282 to 533 eV) through high-order harmonic generation are normally based on inefficient, multi-stage optical parametric amplifiers or optical parametric chirped pulse amplifiers pumped by femtosecond or picosecond lasers. Here we report a very efficient single amplification stage laser based on traditional chirped pulse amplification capable of producing 3 mJ, near-transform limited 44 fs (<6 cycles), 1 kHz pulses centered at 2.5 $\mu$m. The $\approx 68$ GW peak power is the highest value ever reached at this wavelength. In order to fully compress the laser pulses our system is built in a nitrogen box. Our  system utilizes water cooled chromium doped zinc selenide (Cr$^{2+}$:ZnSe) as the gain medium and is pumped by a commercial nanosecond holmium doped yttrium-aluminum-garnet (Ho:YAG) laser.       
\end{abstract}

\maketitle

%\section{\label{sec:level1}Introduction}
The natural time scale of electron dynamics in atoms, molecules, and condensed matter is on the order of attoseconds. Since the first demonstration of tabletop attosecond light sources from high-order harmonic generation (HHG) in 2001~\cite{PaulScience2001, HentschelNature2001}, titanium doped sapphire (Ti:Sapphire) lasers centered at 800 nm have been the workhorse for attosecond research. These systems have enabled coherent pulses covering the extreme ultraviolet spectral range of 15–150 eV~\cite{ChiniNatPho2014,ChangJOSAB2016}. However, for many applications in ultrafast physics and chemistry it is instrumental to extend attosecond capabilities into and beyond the soft X-ray (SXR) “water window,” i.e., $\geq$~530 eV. High-energy attosecond SXR pulses would enable time-resolved studies of core-level transitions in organic and biological systems containing carbon, nitrogen, oxygen, and several other key elements.

In the past two decades many methods of extending the photon energy have been suggested, such as producing HHG in ions or from core level electrons\cite{BrewczykJPB2001,MilosevicPRA2000b}. However, in 2001 it  was experimentally demonstrated that a more straightforward method of extending the cutoff frequency was merely to increase the driving laser wavelength, since the electron quiver energy, and thereby the photon cutoff energy, scales quadratically with the driving wavelength~\cite{shanPRA2001}. In the initial work this was performed with a home-built 100 $\mu$J optical parametric amplifier (OPA) pumped by a few mJ femtosecond Ti:Sapphire laser that was based on traditional chirped pulse amplification (CPA) technology~\cite{StricklandOC1985}. The low pulse energy of the OPA meant that the HHG cut-off energy was limited to about 150 eV for a driving wavelength of 1500 nm, while the cutoff was limited to about 40 eV for 800 nm pulses with the same energy.

In recent years much higher energy (1-2 mJ) OPAs typically centered at ~1800 nm has become commercially available. These systems  typically have pulses durations around 40 fs, but by deploying pulse compression techniques such as gas-filled hollow-core fibers followed by  fused silica wedges less than two-cycle pulses with >0.5 mJ can be produced. This has enabled photon energies covering the SXR water window~\cite{SchmidtOE2011,CousinICUP2016,JohnsonSciAdv2018}. An alternative method of generating high-energy short wave infrared pulses based on optical parametric CPA (OPCPA) have led to carrier-envelope phase stable pulses at ~1.7 $\mu$m with pulse energies up to 3 mJ~\cite{IshiiOL2012,YinOL2016}. 
Generating HHG with these two types of novel systems  it has been possible to generate soft x-ray pulses with durations around 50 as~\cite{LiNatComm2017,GaumnitzOE2017}. Very recently, attosecond transient absorption spectroscopy at the Nitrogen K-edge (400 eV) and Titanium L-edge ( 460 eV) has been demonstrated with both OPAs and OPCPAs\cite{BuadesArX2018,SaitoArX2019}. 

Nevertheless, it is crucial to further extend both the HHG cutoff photon energy and flux in order to enable studies of more complicated systems. In order to significantly improve this we need to improve the driving laser systems.  As an example we will review here the three-stage OPCPA in reference~\cite{IshiiOL2012}. This system is centered at 1.7 $\mu$m and is pumped by a 20 mJ, 4 ps, three-stage Ti:Sapphire CPA. The power amplifiers of the CPA are pumped by more than 100 mJ of nanosecond pulses centered at 527 nm from Q-switched Nd:YLF lasers~\cite{YinOL2016}. The energy conversion efficiency of the 527 nm pump lasers to the 1.7 $\mu$m output is rather low, <3\%, which means that the upkeep cost is rather high for this laser. Moreover, the multi-stage configuration of both the Ti:Sapphire laser and the OPCPA makes it difficult to operate on a day-to-day basis. 

Research have indicated that lasers capable of generating isolated attosecond pulses in the water window and beyond must meet the following requirements:  (1) center wavelength > 1.5 $\mu$m, (2) pulse energy > 0.5 mJ, and (3) pulse duration < 2 cycle.
Recent progress in mid-infrared laser technology~\cite{MirovIEEE2018} suggests alternative laser configurations that can potentially meet these requirements. In particular, CPA based on the four-level gain medium \znse{}  seems to be a promising choice~\cite{SlobodchikovCLEO2011,RenOL2018,VasilyevASSL2019}. The broad vibronic emission spectrum of \znse{} is from 2 $\mu$m to 3.3 $\mu$m, which means that it would in principle be possible to build 3-4 cycle lasers based solely on \znse{}. Furthermore, \znse{} can be pumped using Ho:YAG lasers at 2.09 $\mu$m so the quantum efficiency can be as high as 80\% for a \znse{} centered at 2.5 $\mu$m. Moreover, this gain material can be used to develop multi kHz laser system. 

We have recently reported on a three-stage chirped pulse amplifier utilizing this gain crystal with an output of 2.3 mJ and a pulse duration of 88 fs~\cite{MirovIEEE2018}. In this paper, we demonstrate a CPA system that can produce 3 mJ, 44 fs, 1 kHz pulses in a single-stage amplifier.

%\section{Configuration of the single stage \znse{}  CPA laser }
The layout of the \znse{}  laser system is shown in Figure 1. The seed pulses for the system are generated the same way as in the previously reported three-stage amplifier and is therefore only outlined here. A 14-pass Ti:Sapphire CPA system is used to generate 1.5 mJ, 30 fs, 800 nm pulses at a repetition rate of 1 kHz [14]. These pulses are spectrally broadened and compressed in a conventional argon-filled hollow-core fiber setup followed by a set of double-angle chirped mirrors. The output beam is focused into a 0.8 mm thick BIBO crystal cut for type-I phase matching  that allows for Intrapulse Difference Frequency Generation (IDFG)~\cite{YinSciRep2017}. The IDFG spectrum covers the 1.8  $\mu$m to 4.2 $\mu$m range. Since the emission wavelength of \znse{}  is around 2.5  $\mu$m. A long pass filter is added to block wavelengths below 1.6 $\mu$m. The energy of pulse after the filter is about 8  $\mu$J.  
\begin{figure}
\includegraphics[width=0.54\textwidth]{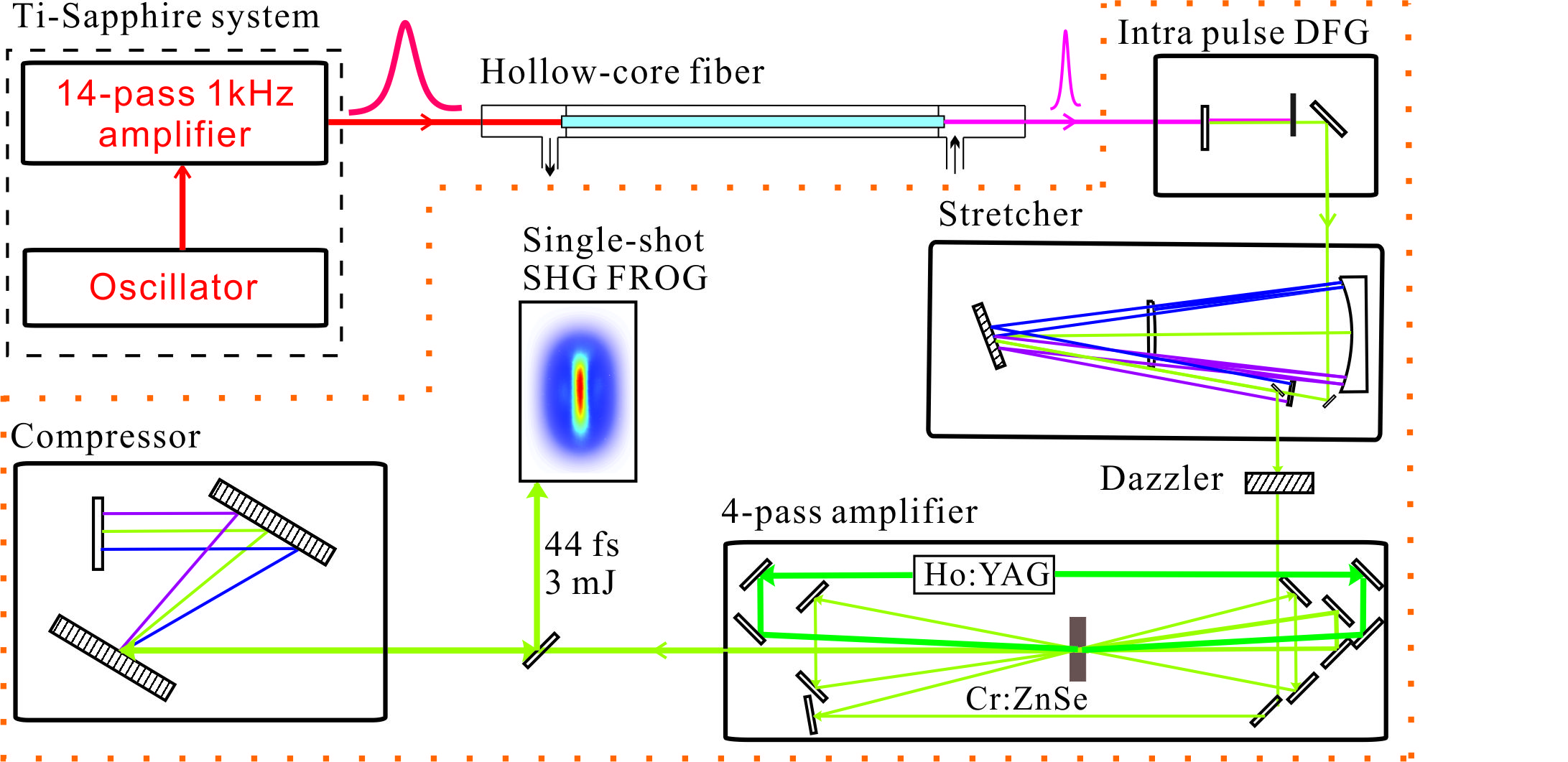}% Here is how to import EPS art
\caption{\label{fig:epsart} Layout of \znse{}  CPA system.}
\end{figure} 

The seed pulse is then sent to an aberration free \"{O}ffner-triplet stretcher containing a reflective grating with groove density of 300 l/mm. The radius of curvature of the convex and the concave mirror of the Öffner telescope is 1 m and 0.5 m, respectively. The stretched pulse duration is about 300 ps. After the stretcher the seed beam has a pulse energy around 3  $\mu$J. A mid-infrared Acousto-Optic Programmable Dispersive Filter, DAZZLER~\cite{MaksimenkaOL2010}, is deployed after the stretcher to pre-compensate high-order phase in the amplifier as detailed below. The DAZZLER further reduces the seed pulses to roughly 1.5 $\mu$J. After the DAZZLER the pulses are directed into a four-pass amplifier. 
The gain medium is a 40 mm long rectangular poly-crystalline \znse{}  crystal with a height of 5 mm and a width of 10 mm. The crystal is water cooled to 18$^\circ$C, and is pumped by a 25 W, 2.09 $\mu$m pump laser (IPG Photonics) from both ends. The diameter of the pump beam is 3.2 mm. After the first two passes the pulse energy is about 20  $\mu$J and 300 $\mu$J, respectively. The amplification factor is therefore $\approx$15 in both cases. The gain starts to decrease after the third pass resulting in an energy of 2.4 mJ, representing an amplification factor around 8. After the fourth pass the gain is saturated with the pulse energy 5.5 mJ. In each pass a lens is placed so that the seed beam is focused before reaching the crystal in order to balance the thermal lensing inside the crystal~\cite{KoechnerAO1970}. This single-stage configuration thus achieves a similar output energy to the recently reported 3-stage \znse{}  amplifier, and we attribute this to a better understanding and management of thermal lensing.  
After the amplifier the pulses are compressed in a double-pass grating compressor with same groove density as the stretcher. The throughput of the compressor is 66\% resulting in an output pulse energy of 3\,mJ. Finally, the temporal profile of the output was characterized by a home-built single-shot second harmonic Frequency Resolved Optical Gating (FROG)\cite{PalaniyappanRevSciInst2010}. 
The entire system is built inside a sealed box, and is purged with nitrogen during operation. The purging is needed, due to strong absorption in water vapour for wavelengths $\geq 2.5 \mu{}$m. 

%\section{Thermal lensing in \znse}
In order to evaluate the effects of thermal lensing of the amplifier we performed simulations using the LAS-CAD software package\footnote{LAS-CAD GmbH Munich, Germany, http://www.las-cad.com/}. Our simulations consisted of finite element analysis of the thermal distribution and lensing of the crystal. The material parameters used for our simulations are listed in Table 1, where $\kappa$ is thermal conductivity, $w$ is the pump beam radius, $P$ is the absorbed average pump power, $\eta$ is the quantum efficiency given by the ratio between the absorption and emission photon wavelengths, $\alpha$ is the linear thermal expansion coefficient, $n$ is the refractive index at the seed wavelength of 2500 nm,  $dn/dT$ is the temperature dependence of the refractive index at this wavelength, $Y$ is Young's Modulus, and finally $\nu$ is Possion’s ratio.
 \begin{table}
\caption{\label{tab:ZnSe}Material properties of Zinc Selenide\cite{Weber2002,SorokinaOM2004}.}
%\begin{ruledtabular}
\begin{tabular}{|ccccc|} 
\hline \rule{0pt}{3ex}  
$\kappa$ [W/(W$\cdot{}$m)] & $w [mm]$ & $P [W]$ &  $\eta [\%]$  &  $\alpha$  [K$^{-1}$]  \\
\hline  \rule{0pt}{2ex}  
$18$ & $1.3$ & $20$ &  $83.6$  &  $7.3\cdot{}10^{-6}$ \\  \hline \rule{0pt}{3ex}  
n& $dn/dT$ [K$^{-1}$] &  $Y$ [GPa]& $\nu$  &\\ \hline  \rule{0pt}{2ex} 
2.44 &$7\cdot{}10^{-5}$ &  $75.8$ & $0.28$ &\\
\hline
\end{tabular}
%\end{ruledtabular}
\end{table}
\begin{figure}
\includegraphics[width=0.49\textwidth]{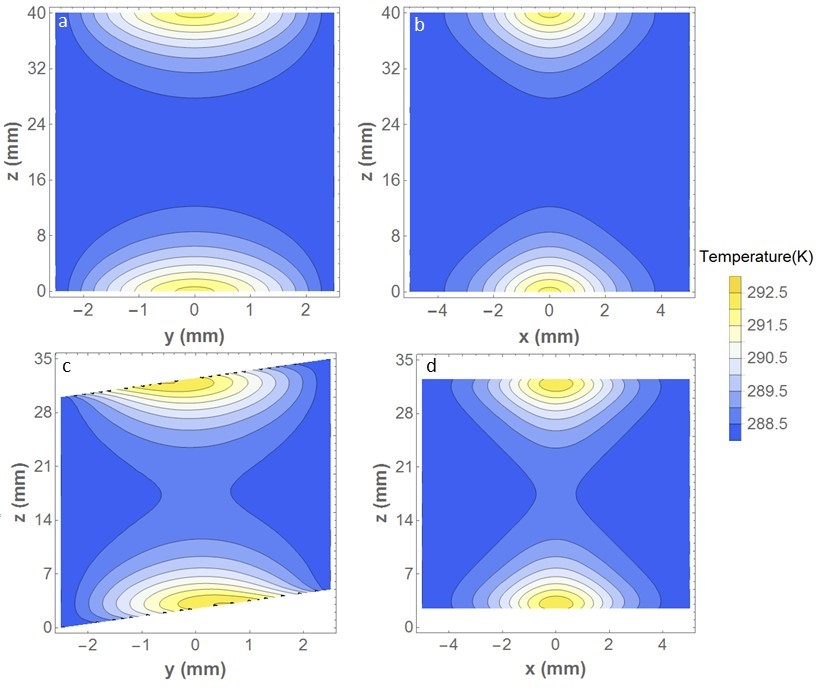}% Here is how to import EPS art
\caption{\label{fig:thermal} Comparison of thermal distribution for crystal for the parameters given in table~\ref{tab:ZnSe} cut for normal incidence (a,b) and Brewster cut crystal (c,d). In all the subfigures x, y and z are the horizontal, vertical and propagation directions, respectively.}
\end{figure} 
In our simulations we compared Brewster cut crystals with plane incidence crystals. Figure \ref{fig:thermal} (a) and 2 (b) shows the simulated temperature distributions for the normal incidence crystal, while (c) and (d) shows this for the Brewster’s cut crystal. In the figures the x, y and z axis represent the horizontal, vertical and propagation directions of the various crystals, respectively. For the normal incidence crystal a minor level of asymmetry between horizontal and vertical axis is observed. Using the software package, we were able to estimate the focal length of the normal incidence crystal to be 1.22 m in the x-z plane and 0.95 m in the y-z plane. The Brewster cut crystal clearly exhibits a higher level of temperature asymmetry between the vertical and horizontal direction. This asymmetry leads to an enhanced amount of astigmatism and other aberrations of the amplified beam. 
 Unfortunately, the software was unable to estimate the aberrations caused by thermal lensing in the Brewster cut crystal, due to the severity of the asymmetry of the temperature distribution.  
Therefore, we attribute the large difference in temperature dependent astigmatism as the reason for the main difference in gain and efficiency between the two laser designs.

\begin{figure}[ht]
\hspace*{-0.25cm}  \includegraphics[width=0.5\textwidth]{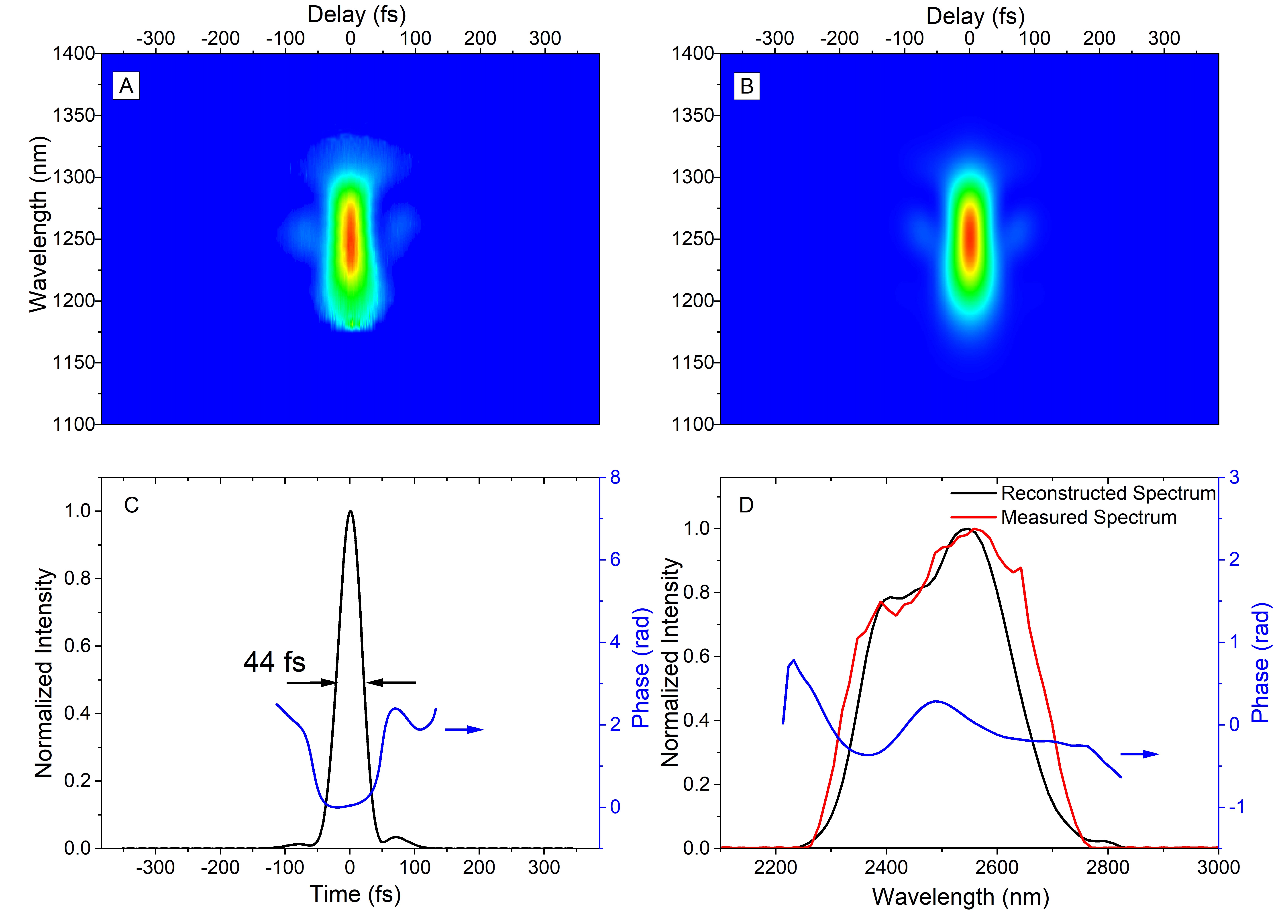}% Here is how to import EPS art
\caption{\label{fig:FROG} FROG measurement and retrieval. (A) Measured FROG
trace. (B) Retrieved FROG trace.  (C) Black curve,
retrieved pulse intensity profile; blue curve, retrieved temporal
phase. (D) Red curve, measured spectrum
with SM301-EX (Spectral Product, LLC); black curve, retrieved
spectrum; Blue curve, retrieved spectral phase.}
\end{figure}

%\section{Dispersion and compression of the system}
The output spectrum of the laser can be seen in figure~\ref{fig:FROG}~(d). The bandwidth supports pulses down to 43 fs. In order to compress the pulses we calculated the dispersion of the stretcher up to fourth order in the Littrow configuration (angle of incidence of 22$^\textrm{o}$).
We also calculated the effects of material dispersion in the amplifier chain up to fourth order based on the Sellmeier equations for CaF$_2$ and ZnSe\cite{TatianAO1984,MalitsonAO1963}, The calculated group delay dispersion (GDD), the third order dispersion (TOD) and the fourth order dispersion (FOD) for the material and stretcher are given in table~\ref{tab:dispcalc}.

The system has a large amount of material dispersion, due to the long gain crystal. We numerically optimized  the grating angle and separation of our grating compressor to compensate the combined GDD and TOD of the stretcher and amplifier. We found that it was possible to compensate this GDD and TOD with a simple double pass 2-grating compressor with the same groove density as the stretcher. However, by doing so we were left with a FOD of -5.5$\cdot 10^6$ fs$^4$, which would limit the pulse duration to about 160 fs. In principle, shorter pulses could be obtained by tweaking the grating angle and separation and obtain a compromise between GDD, TOD and FOD that would lead to a shorter pulse - but not transform limited pulse.
\begin{table}
\caption{\label{tab:dispcalc}Dispersion calculations for 20 mm of CaF2 and 160 mm ZnSe and the \"{O}ffner stretcher at Littrow configuration with a grating separation of 20 cm.}
\begin{ruledtabular}
\begin{tabular}{cccc}
Optical element&GDD (fs$^2$)& TOD (fs$^3$) & FOD (fs$^4$) \\
\hline \rule{0pt}{3ex}    
Stretcher&  $1.25\cdot 10^6$ & $-6.61\cdot 10^6$ &$ 5.16\cdot 10^7$\\
Material&  $-3.25\cdot 10^4$  & $6.67\cdot 10^4$ & $9.2\cdot 10^3$\\
\end{tabular}
\end{ruledtabular}
\end{table}
We therefore chose to deploy a mid-infared DAZZLER to be able to pre-compensate high-order chirp.
The measurements were conducted when the laser enclosure was purged with dry nitrogen and the humidity is <5\% (This humidity is limited by our detector).  The results are shown in Fig. 3. The FWHM pulse duration is 44 fs, which is the shortest pulse for mJ lasers centered at ~2.5 $\mu$m. Experimentally we found that the pulse duration was optimized by adding a FOD of about $3.5\cdot 10^6$ fs$^4$. We attribute the minor differences in experimental values and theoretical values to experimental imprecision in grating angles and separations and residual water vapour after the purging. 
 
%  \textcolor{red}{\textbf{Skip?}The width of the output spectrum limited by gain narrowing can be estimated by\cite{SiegmanLasers1986}
% \begin{equation}
%     \Delta \lambda_\textrm{out} \approx \Delta \lambda_\textrm{a} \sqrt{\frac{3}{10\log{G\left( \Lambda_\textrm{a} \right) -3}}},
% \end{equation}
% where $\Delta\lambda_\textrm{a}\approx 750 $ nm is the FWHM of the gain profile for \znse. $G\left( \Lambda_\textrm{a} \right)$ is the peak of the gain curve, which is ~2000. The calculated $\Delta \lambda_\textrm{out}$=237 nm, corresponding to a Fourier transform limited duration of 36 fs. }

In summary, we have demonstrated a single-stage  \znse{} chriped pulse amplifer capable of producing 3 mJ, 44 fs laser centered at 2.5  $\mu$m. 
The optical-to-optical conversion efficiency from the 2.09 $\mu$m pump to the 2.5 $\mu$m laser is >15\%, which is much higher than our previously reported OPCPA (< 3\%)\cite{YinOL2016}. The wall-plug efficiency is even higher for the \znse{} laser since the power supply of the Ho:YAG laser consumes 1.3 kW whereas the three Nd:YLF lasers for the OPCPA require 5.7 kW to operate (power for chillers not included). The output energy can be further increased by adding more passes and additional amplification stages. The peak power, 68 GW, is the highest achieved at 2.5 $\mu$m. By purging the entire laser system of water vapour with a nitrogen atmosphere and deploying a mid-infrared DAZZLER we were able to halve the pulse duration compared to our previous system\cite{RenOL2018} and produce near transform-limited pulses of 44 fs corresponding to less than 6 optical cycles. Further self-compression to few-cycle pulses in either a solid medium with negative GVD materials~\cite{ZouOL2019} or through hollow-core fiber compression~\cite{SchmidtOE2011} in order to generate isolated attosecond X-rays in the SXR water window seem very feasible. The carrier-envelope phase~(CEP) of the seed pulses to our laser may be stable since they are produced through IDFG. Therefore, stabilizing the CEP of the entire system should be possible by compensating the slow phase drift introduced by the grating based pulse stretcher and compressor by techniques previously developed for Ti:Sapphire lasers~\cite{LiOE2006,FordellOE2009}.

\begin{acknowledgments}
This work was supported by United States Air Force Office of Scientific Research (AFOSR) (FA9550-15-1-0037, FA9550-16-1-0013); Army Research Office (ARO) (W911NF-14-1-0383, W911NF-19-1-0224); Defense Advanced Research Projects Agency (DARPA) (D18AC00011); National Science Foundation (1806575); Defense Threat Reduction Agency (HDTRA11910026). DSTL/EPSRC (MURI EP/N018680/1). We also thank Prof. J. Marangos for his support. 
\end{acknowledgments}

\bibliography{references}% Produces the bibliography via BibTeX.

\end{document}